# Electron–Ion Equilibration in the Merging Galaxy Cluster A665

Christian T. Norseth[1], Daniel R. Wik[1], Craig L. Sarazin[2], Ming Sun[3], and Fabio Gastaldello[4]
[1] Department of Physics & Astronomy, The University of Utah, 115 South 1400 East, Salt Lake City, UT 84112, USA
[2] Department of Astronomy and Virginia Institute for Theoretical Astronomy, University of Virginia, P.O. Box 400325, Charlottesville, VA 22904, USA
[3] Department of Physics & Astronomy, University of Alabama in Huntsville, 301 Sparkman Drive, Huntsville, AL 35899, USA
[4] INAF—IASF Milano, via A. Corti 12, I-20133 Milano, Italy

## Abstract

Galaxy cluster mergers drive powerful shock fronts that heat the intracluster medium (ICM) and accelerate particles, redistributing the energy in a merger. A665 is one of only a few clusters with such a powerful shock ($\mathcal{M} \sim 3$), and it provides a unique opportunity to study the thermalization timescale of the ICM, particularly the electron–ion equilibration timescale. Understanding this timescale is crucial for determining how the energy from the merger is distributed between thermal and nonthermal particle populations. Using $\sim 200$ ks of NuSTAR observations, we measure the temperature distribution across the shock to distinguish between two heating models: (1) an instant collisionless model, where ions and electrons are immediately heated at the shock front; and (2) a collisional model, where electrons are initially adiabatically compressed at the shock and subsequently equilibrate with the ions over $\sim 100$ Myr. Our measurements favor the delayed-equilibration model, suggesting that electrons do not immediately reach thermal equilibrium with the ions at the shock front and instead equilibrate over $t_{\mathrm{eq}} = (4.0 \pm 3.4) \times 10^8$ yr. Additionally, our temperature measurements indicate that the Mach number may be lower than previously estimated ($\mathcal{M} = 2.8 \pm 0.7$), suggesting that the shock strength has been overestimated in past studies. These results add to our understanding of the microphysics governing how thermal energy is distributed in diffuse plasmas like the ICM, with implications for galaxy cluster evolution, large-scale structure formation, and cosmology.

*Unified Astronomy Thesaurus concepts:* Galaxy clusters (584); Intracluster medium (858); Shocks (2086); Plasma astrophysics (1261); Cosmology (343)

## 1. Introduction

Galaxy clusters are the largest gravitationally bound objects in the Universe, comprised of hundreds to thousands of galaxies, with the majority of the baryonic mass residing in a hot intergalactic gas ($T \sim 10^8$ K), known as the intracluster medium (ICM). Bremsstrahlung radiation is the dominant emission mechanism for the ICM, where free electrons pass by heavier ions and emit photons in the X-ray. As the ICM emits X-ray photons, X-ray observations of galaxy cluster mergers can be studied in detail, to provide key insights into cosmology through hydrostatic mass measurements (A. Vikhlinin et al. 2009), particle acceleration mechanisms (A. Sarkar et al. 2024), and ion–electron equilibration timescales (H. R. Russell et al. 2022).

Galaxy clusters grow through mergers of smaller subclusters, and as two clusters collide, the gravitational potential energy of their infalling gas is converted into kinetic energy. As they merge, the gas compresses and the kinetic energy ($\sim 10^{64}$ erg) is converted into thermal energy through the propagation of shocks and turbulence, which heat the ICM (e.g., M. Markevitch et al. 1999). Over time, the shocks and turbulence thermalize the ICM, distributing the thermal energy evenly throughout the gas. The exact timescale for this thermalization is, however, unclear.

There are two widely regarded physically plausible heating scenarios for shock heating: either that ions and electrons are both strongly heated at the shock front; or that only the ions are heated initially and the electrons then equilibrate through collisions with the ions over $\sim 100$ Myr. In the instant model, a strong shock heats both the ion and electron populations as the gas compresses, instantaneously heating both populations. In this scenario, the ions and electrons then reach equilibrium almost immediately. In the collisional model, the gas adiabatically compresses at the shock front as one cluster's ICM moves through the other, driving the electron and ion temperatures up. The ions initially receive more kinetic energy than the electrons, due to their larger mass. Both populations then separately thermalize through collisions among themselves, where the ions reach a higher temperature as they initially received more energy from the shock. The ions and electrons then equilibrate through Coulomb collisions over a timescale of approximately 100 Myr. For a review, see L. Spitzer (1962).

While these scenarios are presented as a binary, current theory does not establish one scenario as being universally applicable over the other. Instead, the dominant heating mechanism—whether through Coulomb scattering (delayed model) or a magnetohydrodynamics process (instant model)—is expected to depend on the properties of the shock, such as the Mach number and the magnetic field configuration in the ICM (J. Vink et al. 2015). For a more in-depth discussion of the models, see Section 4. Given these theoretical uncertainties, observational constraints remain the most reliable way of distinguishing between the models and determining which scenario operates in a given cluster.

Galaxy cluster mergers are ideal places to investigate the two models, due to the hot and diffuse nature of the ICM. Since the Coulomb collisional timescale is strongly dependent on temperature and inversely proportional to density (Equation (4)), and the ICM is very hot and diffuse, this results in a long ($\sim 100$ Myr) equilibration timescale and an observable temperature gradient across the shock, which can be used to distinguish between the two shock-heating models. Measuring this post-shock electron temperature distribution in a cluster relies on specific merger conditions, including a plane-of-sky merger, a sufficiently strong shock, and a large separation between the structures behind the shock and the shock itself.

A plane-of-sky merger is ideal for measuring the temperature gradient and less prone to systematic errors. Since clusters are transparent and the shock can be observed through the ICM, the shock properties can become distorted if the merger axis is not perpendicular to our line of sight. For example, the measured post-shock temperature and subsequent shock strength will decrease as the merger axis tilts farther away from the plane of the sky (U. Chadayammuri et al. 2021). Since the shock will propagate in the plane of the merger, an accurate, undistorted measurement of its properties requires the axis of the merger to be in the plane of sky, as close as possible to a 90° angle from our line of sight.

A strong shock with a high Mach number ($\mathcal{M} > 2$, $\mathcal{M} \sim 3$) is necessary in order to measure statistically significant temperatures. Since the Mach number can be calculated using temperature jumps (Equation (1)), a higher Mach number results from a higher difference in pre-shock and post-shock temperatures. The larger the temperature jump, the less overlap between the temperature measurement uncertainties. In addition, J. Vink et al. (2015) found that the electron and ion temperatures are roughly equivalent at $\mathcal{M} \lesssim 2$, further emphasizing the requirement of a strong ($\mathcal{M} > 2$) shock to evaluate electron–ion equilibration models.

A large separation between the shock and the structures behind the shock is also necessary to accurately measure the impact of the shock. Measurements of the post-shock equilibration area will become distorted if it is overlaid with other ICM structures along our line of sight. The shock must also be not too far into the cluster outskirts, where the ICM is too diffuse to properly measure. Since the emission is proportional to the density squared, the cluster begins to blend into the instrumental background in the outskirts. These unique conditions make identifying a cluster merger and subsequently measuring the post-shock electron temperature a difficult task.

The Bullet Cluster provided the first opportunity to directly measure the electron–ion equilibration timescale and evaluate the heating mechanism. M. Markevitch (2005) favored the instant-shock-heating model in the bow shock ($\mathcal{M} = 3.0 \pm 0.4$), measuring a shorter equilibration timescale than would be predicted by delayed equilibration at 95% confidence ($\sim 2\sigma$). We note that large uncertainties in Chandra measurements of the post-shock electron temperature made it difficult to distinguish between the instant and delayed-equilibration models, based solely on the temperature profile. In a follow-up study, L. Di Mascolo et al. (2019) combined Sunyaev–Zeldovich and X-ray observations to find evidence of delayed equilibration with a lower Mach number of $\mathcal{M} = 2.53^{+0.33}_{-0.25}$ in the bow shock, in good agreement with their X-ray derived Mach number for the same delayed thermalization scenario.

A handful of merging clusters have since had their temperatures behind the shock mapped, in an attempt to constrain the heating mechanism. H. R. Russell et al. (2022) measured the Chandra post-shock electron temperature in A2146's bow shock ($\mathcal{M} = 2.3 \pm 0.2$) and ruled out the instant-shock-heating model with $6\sigma$ confidence, strongly favoring collisional equilibration. This agreed with the previous work of H. R. Russell et al. (2012), which determined that the equilibration timescale was consistent with the delayed-equilibration model in the bow shock and that the post-shock temperature was lower than predicted by instant shock heating, further favoring delayed equilibration. Both the H. R. Russell et al. (2012) and (2022) works favored instant shock heating in the upstream shock ($\mathcal{M} = 1.5 \pm 0.1$) with hot post-shock temperatures, but the temperature values carried large uncertainties and were complicated by substructures.

A. Sarkar et al. (2024) analyzed Chandra observations of A1240 and favored collisional heating in both the southeastern shock ($\mathcal{M} = 1.49^{+0.22}_{-0.24}$) and the northwestern shock ($\mathcal{M} = 1.41^{+0.17}_{-0.19}$), but also stated that the large uncertainties in temperature measurements prevented the ruling out of either model. The measurements of these clusters highlight the statistical uncertainty present when distinguishing between the instant-shock-heating and the delayed-equilibration models.

To make matters more complicated, temperatures measured between the prominent X-ray observatories NuSTAR (F. A. Harrison et al. 2013) and Chandra disagree with each other. Chandra consistently measures hotter temperatures than NuSTAR at higher energies, increasing the uncertainty between the two equilibration models (C. Potter et al. 2023). The discrepancy in the temperature measurements is still under investigation, but a potential calibration solution has been found (see F. Lopez et al. 2025, in preparation). NuSTAR's greater sensitivity at higher energies makes it more sensitive to hotter gases. Since galaxy cluster mergers are extremely energetic, with hot temperatures, NuSTAR is thus appropriately positioned to study them.

NuSTAR observations of A665 provide a unique opportunity to map the post-shock electron temperature and to constrain the heating mechanism. Chandra observations of A665 ($z = 0.1819$, $M_{500} = 8.86 \times 10^{14} M_\odot$) revealed a strong shock ($\mathcal{M} = 3.0 \pm 0.6$), making it one of the few major galaxy cluster mergers detected with such a powerful shock, roughly in the plane of the sky, and with a large separation between the central cluster emission and the shock front (S. Dasadia et al. 2016b; V. Cuciti et al. 2022). It is characterized by a disrupted cool-core remnant from the roughly-equal-in-size infalling subcluster as it passed through the main cluster from the northwest to the southeast (P. L. Gomez et al. 2000; M. Markevitch & A. Vikhlinin 2001; F. Govoni et al. 2004). The merger axis is in the southeast–northwest direction, with a prominent cold front and radio halo in the southeast and a strong shock and radio halo in the northwest (S. Dasadia et al. 2016b). The cluster likely had a nonzero impact parameter, which explains why both core remnants survived. Using follow-up observations with NuSTAR, we map the electron temperature behind the northwest shock front and constrain, possibly for the first time, how the ICM is heated at the shock front.

This work is organized in the following way. Section 2 describes the data used and the data reduction techniques. Section 3 goes through the analysis methods and results. We discuss the results in Section 4 and conclude the work in Section 5. A figure displaying important regions can be found in Figure 1. In this work, we assume a flat $\Lambda$CDM cosmology with $H_0 = 70 \, \mathrm{km \, s^{-1} \, Mpc^{-1}}$, $\Omega_m = 0.27$, and $\Omega_\Lambda = 0.73$. At A665's redshift ($z = 0.1819$), $1''$ corresponds to 3.073 kpc. Unless otherwise specified, the uncertainty ranges are given as 90% confidence intervals.

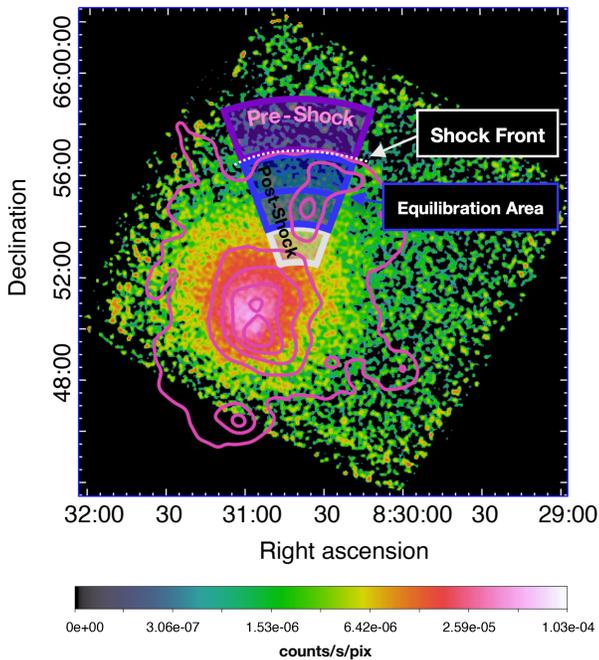

**Figure 1.** An exposure-corrected, background-subtracted, and smoothed 4–25 keV NuSTAR image of A665 with regions overlaid. The four regions correspond to the data points in Figure 4, with the purple region corresponding to the pre-shock area, the two blue regions covering the equilibration area, and the white region corresponding to an additional post-shock region. The shock front is indicated by the dashed white line. LOFAR radio contours are overlaid in pink. The color bar is in units of counts per second per pixel. For more details of the processing and presentation of this image, see Section 2 and Figure 2.

## 2. Observations and Data Reduction

### 2.1. NuSTAR Observations

We use two NuSTAR observations of A665 (Observation IDs: 70201002002 and 70201003002), observed on 2017 May 10 and 12. The first observation had a total exposure time of ∼100 ks and the second had an exposure time of ∼98 ks. See Table 1 for more details. The data were filtered using the standard pipeline processing with HEASoft (v. 6.28) and NuSTARDAS (v. 2.0.0). To avoid periods of high background, event files are typically cleaned automatically, using the SAAMODE = STRICT and TENTACLE = yes flags in nupipeline, which remove periods of high background by excluding data during passages through the South Atlantic Anomaly (SAA) and a "tentacle"-like area near the SAA. Since the automatic filtering processing can be too restrictive and can remove valuable data, we opted to instead manually filter the data, by producing light curves with lcfilter,[5] binned in 100 s time bins. Increased count rates are excluded manually to produce good time intervals (GTIs). This manual filtering is carried out in the 50–160 keV energy range, to exclude high-background counts due to the SAA, and in the 1.6–20 keV energy range, to remove high background associated with solar activity. The new set of filtered GTIs is then reprocessed using nupipeline[6] and images are created using XSELECT in different energy bands, along with exposure maps, using nuexpomap. Then, nuproducts is used to create spectra in specified regions of interest, along with their corresponding Response Matrix Files and Ancillary Response Files (ARFs), to serve as first estimates of the parameter values in each region for later use.

The background was modeled using nuskybgd, which has become a standard approach in the background-dominated regime. The procedures and treatment of the background components are outlined in D. R. Wik et al. (2014). Using spectra extracted from regions roughly coincident with each of the four detectors in the focal plane, with the cluster emission largely excluded, nuskybgd is used to fit the spectra with an a priori model. The regions used in the background analysis and the fits to the spectra with the background model are shown in Figure 6. The background model, defined over the entire focal plane, can then be used to produce background images and background spectra for any region of interest. Using the exposure maps previously produced, a background-subtracted, exposure-corrected image across different energy ranges is created. See the left panels in Figure 2 for background-subtracted, exposure corrected, and smoothed NuSTAR images of A665 in the 4.0–25.0 keV energy range.

### 2.2. Chandra Observations

We refer to Chandra observations of A665 in certain sections of our analysis. The processing of these observations included both a modern and an older version of the CIAO[7] set of analysis tools, the calibration database (CALDB), and the High Energy Astrophysics Software (HEASoft).[8]

The older version of the calibration files and software included CIAO v4.7, CALDB v4.6.9, and HEASoft v6.16. We followed the same data processing technique used in S. Dasadia et al. (2016b), which followed S. Dasadia et al. (2016a) for stowed background handling and subsequent modeling. This included reviving the old CIAO, CALDB, and HEASoft files used in S. Dasadia et al. (2016b) and using them in a virtual machine running an outdated Linux operating system, CentOS5, with packages set up from the available archives.[9] The purpose was to recreate the original analysis in S. Dasadia et al. (2016b) for direct comparison to our NuSTAR results and to a modern Chandra analysis.

We processed the same Chandra observations using a modern version of the calibration files and software, which included CIAO v4.12, CALDB v4.9.1, and HEASoft v6.30.1. For this analysis, we used acisgenie,[10] following the same approach as in C. Potter et al. (2023).

## 3. Data Analysis and Results

With the cleaned data products produced and the background characterized, spectra from regions of interest were extracted and corresponding background spectra (constructed from the full field-of-view background model using nuskybgd) were produced.

### 3.1. Initial Temperature Map Construction

First, the cluster was broken up into regions, including regions around the shock front, to isolate regions of roughly constant temperature. Following the same approach as

---

[5] https://github.com/danielrwik/reduc
[6] https://heasarc.gsfc.nasa.gov/docs/nustar/analysis/nustar
[7] https://cxc.cfa.harvard.edu/ciao/
[8] https://heasarc.gsfc.nasa.gov/lheasoft/
[9] https://vault.centos.org/5.11/
[10] https://gitlab.com/qwq/acisgenie

Table 1
Observations Used in Our Analysis

| Observation ID | Date (yyyy-mm-dd) | R.A. (hh mm ss) | Decl. (deg arcmin arcsec) | Exposure Time (ks) |
|---|---|---|---|---|
| NuSTAR: 70201002002 | 2017-05-10 | 08 30 22.0 | +65 53 34 | 99.9 |
| NuSTAR: 70201003002 | 2017-05-12 | 08 30 26.1 | +65 51 23 | 98.2 |
| Chandra: 12286 | 2011-01-09 | 08 30 59.25 | +65 50 26 | 47.1 |
| Chandra: 13201 | 2011-01-06 | 08 30 59.25 | +65 50 26 | 48.7 |
| Chandra: 3586 | 2002-12-28 | 08 30 53.30 | +65 50 2.4 | 29.7 |

**Note.** R.A. and decl. refer to the pointing position.

C. Potter et al. (2023), an initial temperature map for OBSID 70201002002 was produced by spectral fitting predefined pixel areas. To accomplish this, first narrowband images are made. These include background-subtracted, exposure-corrected images in the 3–5 keV, 6–10 keV, 3–20 keV, and 10–20 keV bands, using nuproducts and nuskybgd, as outlined in Section 2. Circular regions with a radius of 5 pixels are then drawn every 5 pixels across the images. The region size is increased from the minimum radius of 5 pixels until 500 counts are encompassed. This is carried out for each of the three bands, including the raw, background, and exposure maps. The result is a set of coarse spectra binned from the exposure-corrected, background-subtracted counts in each of the bands in each of the circular regions. These spectra were then fit with a single apec model, to find the best-fit temperature in each region, where the abundance and redshift were fixed to values obtained from a global fit to the cluster. To create a temperature map, the best-fit temperature in a region was assigned to the corresponding central pixel, and the temperature between each central pixel was interpolated, to produce the resultant map. The interpolated temperature map can be seen in Figure 7. It is important to note that this is only a rough first analysis used to search for any prominent temperature fluctuations in the ICM.

### 3.2. Region Selection

Using the basic temperature map as a starting point, regions of similar temperature were identified. Surface brightness images of the cluster are used in conjunction with the basic temperature map to select regions that cover the entire cluster.

The sizes and placements of the regions were varied, in order to achieve a high enough signal-to-noise ratio (S/N) for the nuCrossARF analysis to be successful. This included a width of at least 30″ across for each region, ensuring that they were not dominated by scattered emission due to NuSTAR's point-spread function (PSF).

To achieve a high enough S/N, the equilibration area is broken into only two regions, with the shock front boundary identified as a discontinuity in the surface brightness profile for the cluster. To locate this discontinuity, we extracted a surface brightness profile from a processed Chandra observation of A665, which was used due to Chandra's superior spatial resolution and to be consistent with the previous work in S. Dasadia et al. (2016b). The shock edge was also compared to LOFAR[11] radio data of A665 taken from LoTSS DR2 (A. Botteon et al. 2022), which confirmed the shock-edge location and demonstrated a shock–radio halo connection, where merger-driven shocks in galaxy clusters can accelerate particles or reaccelerate existing cosmic-ray electrons, which then emit synchrotron radiation. A665 is one of a small handful of clusters with such an association (R. J. Weeren et al. 2019).

We further varied the size and location of the region near the shock front, to assess the impact on our NuSTAR spectral analysis. By using a high-resolution Chandra image for direct comparison to the surface brightness discontinuities, we found that shifting the region south of the shock front by up to 20″ and stretching it by up to twice its original size yielded post-shock temperature measurements that all fell within their respective 90% confidence intervals.

Between all of the region modifications, the post-shock temperatures had a maximum variation of 1.4 keV (18%) from the mean, which was only 40% of the average combined 90% confidence error bars, suggesting that each measurement was consistent within statistical uncertainties. We therefore adopted a post-shock region that closely matched the region used in S. Dasadia et al. (2016b), for easy comparison to previous results, which fell well within our tested parameter space.

While the region selection was guided by Chandra images, this approach was well justified, given Chandra's superior spatial resolution for identifying shock morphology and NuSTAR's superior sensitivity at higher energies. The temperature in the post-shock region lies near the upper end of Chandra's bandpass, where its spectral constraints become weaker. NuSTAR, by contrast, retains high sensitivity at and above 10 keV, making it especially well suited for measuring such high temperatures. Therefore, even though the region was defined using Chandra data, it remained physically appropriate and statistically robust for our NuSTAR spectral analysis.

A comparison of the equilibration region that sits along the shock edge to the LOFAR image of A665 can be found in Figure 3. The final region selections for all of the regions can be seen in the top left panel of Figure 2.

### 3.3. Accounting for Crosstalk between Regions

Due to NuSTAR's large PSF—FWHM ∼ 18″ and half-power diameter ∼ 1′—photons originating from one region are scattered into surrounding regions, and vice versa, causing cross-contamination between regions. nuCrossARF accounts for this by producing cross-ARFS: ARFs for sources of emission that lie outside the region of interest. This goes beyond the standard ARF generation in nuproducts, which only produces ARFs for emission inside of a region of interest. Using the regions selected in Section 3.1, the nuCrossARF code is used to account for scattered light throughout the cluster and to fit the spectra for each region, to isolate each region's true temperature. We use abundance and redshift values obtained from a global fit to the cluster, changing the redshift from 0.1819 to 0.1947. See Appendix B of R. R. Bolivar et al. (2023) for an explanation of why leaving the redshift free in the global fit and using the result in the subsequent analysis gives a better fit. The results and 90% confidence intervals from this simultaneous spectral fitting are shown in Figure 8, with the temperature values given in Table 2 and spatially mapped in the right panels of Figure 2.

---
[11] https://lofar-surveys.org/planck_dr2.html

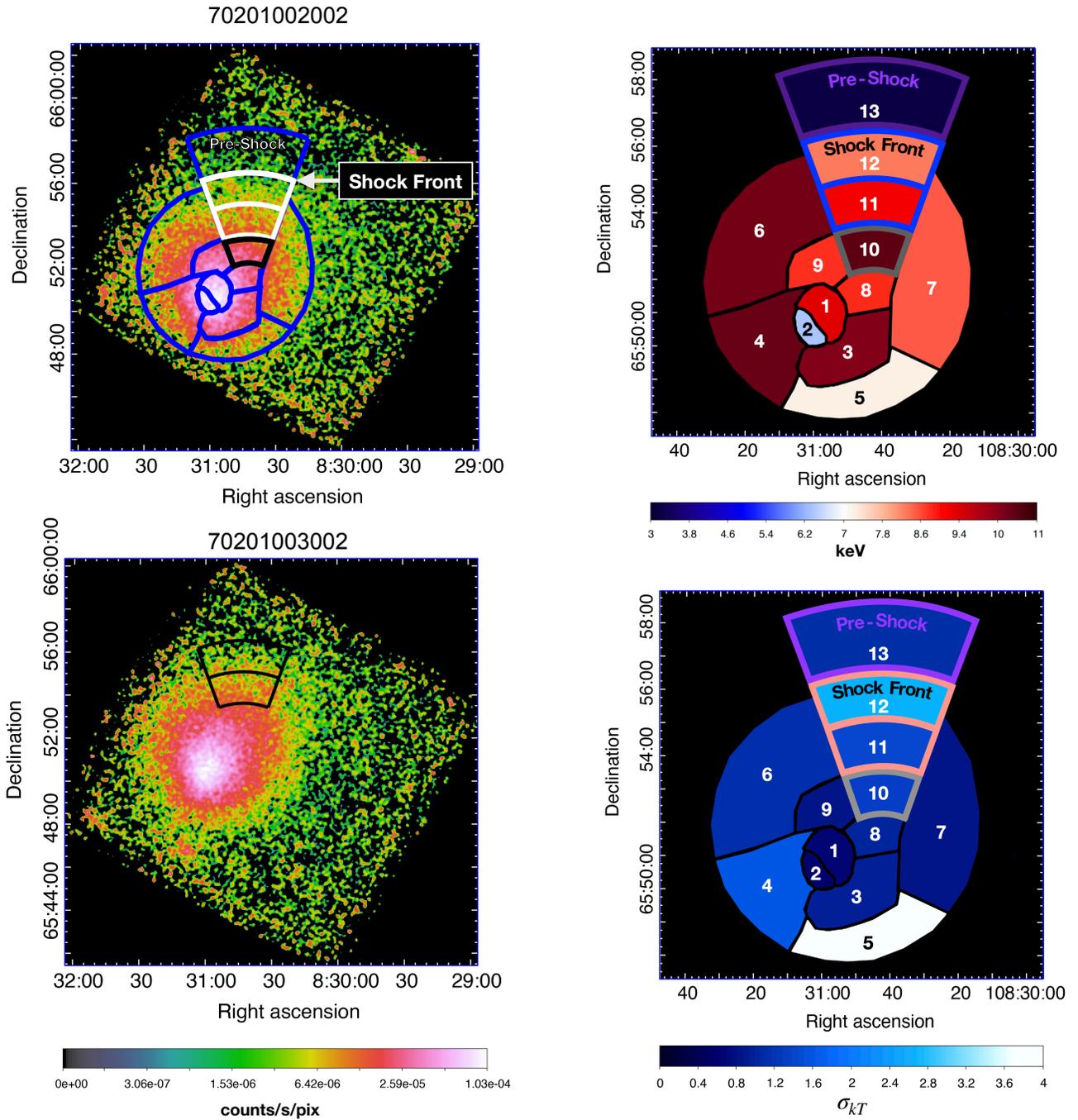

**Figure 2.** Top left: regions chosen for the nuCrossARF analysis overlaid on an exposure-corrected, background-subtracted 4–25 keV image of A665 (OBSID 70201002002). The white regions cover the equilibration area. The pre-shock region and shock front are also labeled for clarity. Bottom left: exposure-corrected, background-subtracted 4–25 keV image of A665 (OBSID 70201003002). The black regions correspond to the equilibration area. Both images on the left are smoothed with a Gaussian kernel with an FWHM of 3 pixels, scaled logarithmically from 0 counts s$^{-1}$ pix$^{-1}$ (black) to 10+ counts s$^{-1}$ pix$^{-1}$ (white), have a pixel size of 2″.46, and have the same color bar in units of counts s$^{-1}$ pix$^{-1}$. Top right: temperature map displaying results from the nuCrossARF analysis, corresponding to the temperatures in Table 2, with an average error map shown below. The regions highlighted in blue cover the equilibration area. The color bar indicates the temperature in keV. The blue and gray outlined regions correspond to the data points in Figure 4. Bottom right: temperature error map showing the average error for each region's temperature measurement in the nuCrossARF analysis, corresponding to the above temperature map. The regions highlighted in cyan indicate the equilibration area. The color bar indicates the average 90% uncertainty, in keV. The equilibration area (Regions 11 and 12), along with an additional post-shock region (Region 10) and the pre-shock region (Region 13), as highlighted in each of the panels, correspond to the four data points displayed in each panel of Figure 4.

Note that the temperature of the outermost region beyond the shock front (Region 13) is not strongly constrained by our NuSTAR data; we have adopted the temperature of 3.2 keV determined by Chandra instead (S. Dasadia et al. 2016b). We further note that when pre-shock Region 13 is removed entirely from the nuCrossARF analysis, the temperature measurement in the post-shock Region 12, which would contain the most scatter from Region 13, only changes by 0.23%, significantly below the uncertainty of the measurement itself (∼30%). However, we still include the pre-shock region in the nuCrossARF analysis, for completeness.

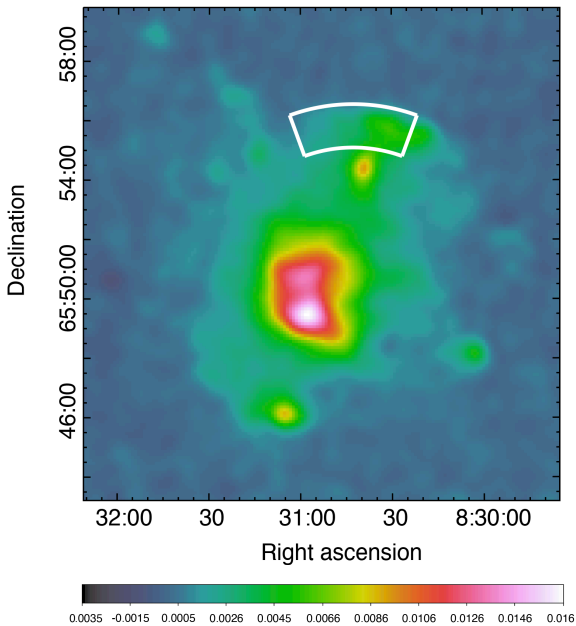

**Figure 3.** The equilibration region that sits along the shock edge used in our analysis (known as Region 12) compared to a LOFAR image of A665, with data tapered to 100 kpc resolution and compact sources subtracted ($\sigma_m = 228\,\mu$Jy/beam), taken from LoTSS DR2 (A. Botteon et al. 2022). The color bar is in units of janskys per beam. The edge of the chosen shock region lies directly along the edge of the shock, as highlighted by the synchrotron radio emission due to particles accelerated by the shock, reaccelerated cosmic-ray electrons compressed by the shock, and subsequently reaccelerated electrons due to turbulence from the merger.

### 3.4. Deprojection

In order to properly compare the temperatures behind the shock to different 3D heating models, the temperatures in the relevant regions need to be deprojected. Since the ICM is transparent, an observation of a galaxy cluster is a projection of a spherical volume of gas onto a 2D plane. Spectra extracted from a particular radial bin are thus projections of the gas through the entire cluster at that radius and at greater radii. We use the XSpec projct[12] model to deproject the temperature fits from nuCrossArf (Section 3.3) in three regions of interest: an inner and outer region in the equilibration area and an adjacent region located farther into the cluster, behind the shock and equilibration region. A pre-shock region fixed at 3.2 keV is also included in the deprojection but has little to no effect on the deprojection results.

The XSpec projct model determines 3D model parameters using 2D projected spectra obtained from observations, by projecting spherical shells (and their corresponding models) into 2D elliptical annuli. Because the routine fits the data directly, PSF-scattered light from all other regions in the cluster had to be included in the deprojection as well. Using

---

[12] https://heasarc.gsfc.nasa.gov/xanadu/xspec/manual/Models.html

**Table 2**
Results from the nuCrossARF Analysis (Section 3.3)

| Region | $kT$ (keV) | Norm[a] ($10^{-3}$ cm$^{-3}$) |
|---|---|---|
| 1 | $9.33^{+0.70}_{-0.60}$ | $2.33^{+0.12}_{-0.12}$ |
| 2 | $6.32^{+0.62}_{-0.62}$ | $1.90^{+0.20}_{-0.16}$ |
| 3 | $10.20^{+1.02}_{-0.92}$ | $1.34^{+0.09}_{-0.08}$ |
| 4 | $10.46^{+1.66}_{-1.55}$ | $0.91^{+0.10}_{-0.08}$ |
| 5 | $7.19^{+5.44}_{-2.58}$ | $0.24^{+0.11}_{-0.08}$ |
| 6 | $10.39^{+1.26}_{-1.00}$ | $1.25^{+0.08}_{-0.08}$ |
| 7 | $8.49^{+0.92}_{-0.76}$ | $1.14^{+0.08}_{-0.08}$ |
| 8 | $8.83^{+1.14}_{-0.89}$ | $0.94^{+0.08}_{-0.08}$ |
| 9 | $8.74^{+0.92}_{-0.76}$ | $1.17^{+0.09}_{-0.08}$ |
| 10 | $10.60^{+1.39}_{-1.25}$ | $0.71^{+0.06}_{-0.05}$ |
| 11 | $9.14^{+1.57}_{-1.29}$ | $0.52^{+0.06}_{-0.05}$ |
| 12 | $8.28^{+3.29}_{-2.06}$ | $0.25^{+0.06}_{-0.05}$ |
| 13 | $3.20$[b] | $0.01$[b] |

**Note.** The solar abundance[c] and redshift parameters in the apec model were fixed in the spectral fitting to 0.3 and 0.1947, respectively. The region numbers correspond to the ordering in Figure 2.
[a] The normalization parameter in the apec model is given as $\frac{10^{-14}}{4\pi[D_A(1+z)]^2}\int n_e n_H dV$.
[b] The Region 13 values are fixed, due to low NuSTAR counts, and are instead taken from Chandra measurements (S. Dasadia et al. 2016b).
[c] The default angr abundance table in xspec was used (E. Anders & N. Grevesse 1989) in the fitting.

the nuCrossARF-generated ARF files and the best-fit models determined in Section 3.3, we simulate fake spectra, using XSpec's fakeit command to simulate the emission contributed from each region into each of the regions of interest. The simulated spectra are added to the background spectra used in the deprojection fitting for each relevant region. The results of the deprojection compared to the equilibration models (see Section 4 for details) can be seen in Figure 4.

## 4. Discussion

### 4.1. Electron–Ion Equilibration

Galaxy cluster mergers release an enormous amount of energy through shocks and turbulence, which heat the ICM. As one cluster's ICM passes through the other, the gravitational potential energy of the gas is converted into kinetic energy, and as the two clusters merge, pressure builds up ahead of the infalling gas as it compresses. Shocks form, since the collision speed is mildly supersonic, which convert the kinetic energy into thermal energy and redistribute the total energy in the merger. The timescale for the deposition of the kinetic energy into the electron population, which produces the Bremsstrahlung emission observed, is unclear.

We compare two heating models in this work: an instant-equilibration model and a Coulomb collisional model.

#### 4.1.1. Instant-equilibration Model

In the case of instant collisionless equilibration, electrons and ions are simultaneously heated at the shock front. The shock imparts energy to both populations, rapidly increasing their temperatures. The electrons and ions then thermalize within their respective populations, each reaching a Maxwellian distribution. The thermal equilibration timescale for the more massive protons is approximately $\sqrt{m_p/m_e} \approx 43$ times slower than for the electrons. However, if both populations receive energy from the shock simultaneously, no equilibration between ions and electrons needs to take place—rendering collisional electron–ion equilibration unnecessary—effectively happening instantly.

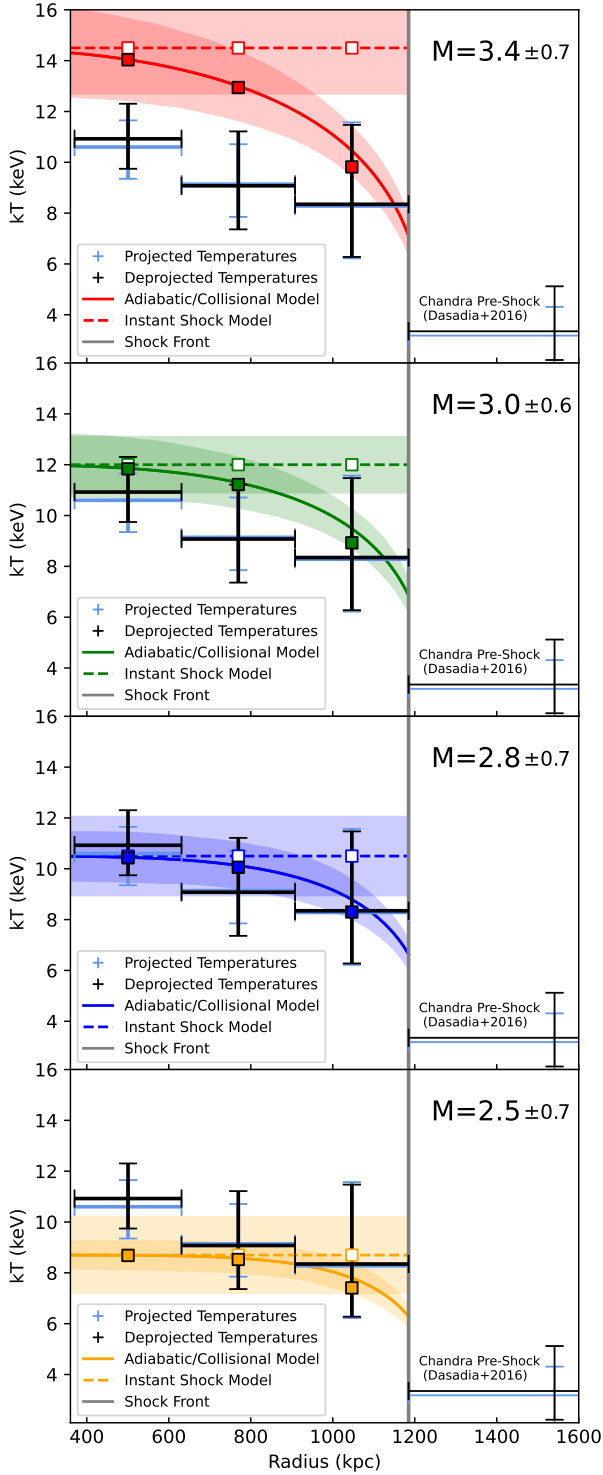

Figure 4. Projected (blue) and deprojected (black) NuSTAR temperatures in the equilibration regions compared to collisional (solid) and collisionless (dashed) models. The colored squares signify the weighted average for each model in each region. The shock front is indicated by the solid gray vertical line. The Chandra pre-shock temperature is also included. The shaded areas represent the $1\sigma$ uncertainties for each model. The red models used a Mach number of $\mathcal{M} = 3.4 \pm 0.7$ (the measured Chandra density jump), the green models used $\mathcal{M} = 3.0 \pm 0.6$ (the Chandra temperature jump), the blue models used $\mathcal{M} = 2.8 \pm 0.7$ (the predicted NuSTAR temperature jump from Equation (7)), and the orange models used $\mathcal{M} = 2.5 \pm 0.7$ (the observed NuSTAR post-shock + Chandra pre-shock temperature jump without accounting for cross-calibration differences).

In the instant-heating model, the post-shock electron temperature follows directly from the Rankine–Hugoniot jump conditions:

$$\frac{T_{\rm post}}{T_{\rm pre}} = \frac{[(\gamma - 1)\mathcal{M}^2 + 2][2\gamma\mathcal{M}^2 - (\gamma - 1)]}{(\gamma + 1)^2 \mathcal{M}^2}, \quad (1)$$

where $T_{\rm post}$ is the post-shock electron temperature, $T_{\rm pre}$ is the pre-shock electron temperature, $\mathcal{M}$ is the Mach number, and $\gamma = 5/3$ is the adiabatic index for a monotonic gas.

### 4.1.2. Collisional Equilibration Model

In the Coulomb collisional model, electrons and ions are initially both heated through adiabatic compression at the shock, where the post-shock electron temperature is determined using

$$T_{\rm post} = T_{\rm pre}\left(\frac{\rho_{\rm post}}{\rho_{\rm pre}}\right)^{\gamma - 1}, \quad (2)$$

where $\rho_{\rm post}/\rho_{\rm pre}$ is the ratio between the pre-shock and post-shock gas densities.

The more massive ions initially receive a greater fraction of the kinetic energy from the shock, driving their temperature to be higher than the electrons. Both the ions and electrons then thermalize among their respective populations to each reach a Maxwellian distribution.

The ions and electrons then equilibrate with each other through collisions at a rate governed by

$$\frac{dT_e}{dt} = \frac{T_i - T_e}{t_{eq}}, \quad (3)$$

where $T_i$ is the ion temperature, $T_e$ is the electron temperature, and $t_{\rm eq}$ is the equilibration timescale for Coulomb scattering given by L. Spitzer (1962) and K.-W. Wong & C. L. Sarazin (2009):

$$t_{\rm eq}(e, p) \approx 2.33 \times 10^8 \, {\rm yr} \left(\frac{T_e}{10^8 \, {\rm K}}\right)^{3/2}$$
$$\times \left(\frac{n_e}{10^{-3} \, {\rm cm}^{-3}}\right)^{-1}, \quad (4)$$

where $n_e$ is the electron density.

As they collide, the ion temperature will decrease and the electron temperature will increase, to eventually equilibrate and reach a mean gas temperature $T_{\rm gas}$, given by

$$T_{\rm gas} = \frac{n_e T_e + n_i T_i}{n_{\rm gas}} = \frac{1.1 T_e + T_i}{2.1}, \quad (5)$$

where $n_i$ is the ion density, $n_e = 1.1 n_i$, $n_{\rm gas} = n_e + n_i$, and $T_{\rm gas}$ is constant with time, representing the final equilibrium temperature when $T_e = T_i = T_{\rm gas}$.

In summary, we calculate the post-shock electron temperature (the temperature immediately behind the shock) using Equation (2), to find the adiabatically compressed electron temperature. For A665, this corresponds to a post-shock electron temperature of $T_{\rm post} = 7.2 \pm 4.2$ keV. The Rankine–Hugoniot jump conditions (Equation (1)) in combination with the pre-shock electron temperature are used to calculate the average gas temperature $T_{\rm gas}$, which is the temperature the electrons will heat to and the ions will cool down to as they equilibrate. For A665, this corresponds to an equilibrium temperature of $T_{\rm gas} = 12.0 \pm 1.1$ keV. The post-shock ion temperature is determined with Equation (5), rearranged to find the ion temperature using the gas temperature and the post-shock electron temperature. For A665, this corresponds to a post-shock ion temperature of $T_{\rm i} = 17.3 \pm 5.1$ keV. The ions and electrons then equilibrate over a timescale set by Equation (4). For A665, this corresponds to an equilibration timescale of $t_{\rm eq} = (4.0 \pm 3.4) \times 10^8$ yr. Equation (3) can be solved analytically to find the evolution of the electron temperature as a function of distance behind the shock front. For details of the solution and the final model used, see Appendix B.

We also calculate the post-shock gas velocity as part of the final model calculations. For A665, this corresponds to a pre-shock sound speed of $c_{\rm s} = (9.3 \pm 1.6) \times 10^2$ km s$^{-1}$, a shock speed of $v_{\rm shock} = (2.8 \pm 0.2) \times 10^3$ km s$^{-1}$, and a post-shock gas velocity of $v_{\rm ps} = (8.8 \pm 4.2) \times 10^2$ km s$^{-1}$.

Since each temperature measurement is an average within the radial range for each region, we also calculate the weighted average for the model within each region, weighted by the post-shock electron density squared. The post-shock electron density is represented by a power-law fit to the Chandra surface brightness profile: $n_{\rm e}(r) = n_{e0} r^{-p}$, where $p = 1.1 \pm 0.1$ and $n_{e0}$ is a normalization factor equal to 1.1 (S. Dasadia et al. 2016b).

We used the `uncertainties`[13] Python package to propagate $1\sigma$ measurement uncertainties in intermediate calculations. For the final model calculations, which required inverting the function (see Appendix B), we adopted a Monte Carlo method to capture the resulting uncertainties in the transformed quantities. For clarity in the presentation, we smoothed the uncertainty bands using a Savitzky–Golay filter, to reduce visual noise while preserving the overall trend (A. Savitzky & M. J. E. Golay 1964). The final models with $1\sigma$ uncertainty bands and weighted averages are compared to the measured NuSTAR temperatures in Figure 4.

### 4.2. Systematic Uncertainties/Mach Number

Our measurements of the post-shock electron temperatures shown in Figure 4 favor the adiabatically compressed collisional model over the instant collisionless model. However, the instant model cannot be unequivocally ruled out, due to the large error bars in our temperature measurements. Furthermore, there is uncertainty in the derived Mach number used to calculate the two models.

The Mach number can be derived in two ways: from the temperature jump at the shock front (via Equation (1)) or from the density jump at the shock front, via

$$\frac{\rho_{\rm post}}{\rho_{\rm pre}} = \frac{(\gamma - 1)\mathcal{M}^2}{(\gamma - 1)\mathcal{M}^2 + 2}, \quad (6)$$

where the terms are the same as previously defined.

The shock-front boundary is defined by a discontinuity in the density profile of the cluster, where the pre-shock and post-shock densities are the density values at the discontinuity and define the density jump. The density profile for a cluster is derived from fitting the cluster's surface brightness profile to a density model. The density model relies heavily on the assumption of spherical symmetry and is significantly impacted by the orientation of the merger with respect to the plane of the sky and our line of sight. For example, as the merger axis tilts away from the plane of the sky, the shock intensity appears to decrease (U. Chadayammuri et al. 2021). These various factors introduce uncertainty in the derived density profile and the measured density jump, which impacts the resultant Mach number. As an example, S. Dasadia et al. (2016b) measure A665's density jump at the shock front with Chandra to be $\rho_{\rm post}/\rho_{\rm pre} = 3.2 \pm 0.3$, resulting in a Mach number of $\mathcal{M} = 3.4 \pm 0.7$. However, their density models include assumptions based on the curve of the shock, which can be altered to assume a greater radial distance, which instead results in a density jump of $\rho_{\rm post}/\rho_{\rm pre} = 2.6$ and a Mach number of $\mathcal{M} = 2.5$. We note the authors did not state the errors on the density jump or Mach number in this alternative scenario.

Calculating the Mach number from the pre- and post-shock temperatures and the subsequent temperature jump relies on spectral analysis, which is less susceptible to uncertainty from geometric and projection effects. Extracting and fitting spectra from a defined region provide a direct measurement of the thermal state of the gas in that region. Therefore, using a temperature jump to measure the Mach number for the shock may be more accurate, if typically less precise.

S. Dasadia et al. (2016b) measure a temperature jump at the shock front using Chandra of $T_{\rm e,post}/T_{\rm e,pre} = 3.8 \pm 1.3$, corresponding to a Mach number of $\mathcal{M} = 3.0 \pm 0.6$. However, relying on temperature measurements from the Chandra telescope introduces its own uncertainty to potential calibration issues. A discrepancy between the measured temperatures from Chandra and NuSTAR has been identified in recent years, especially with hotter Chandra temperatures being consistently higher than those of NuSTAR (A. N. Wallbank et al. 2022; C. Potter et al. 2023). Therefore, assuming a Mach number for the shock derived from the S. Dasadia et al. (2016b) temperature measurements is not necessarily more reliable than the Mach number derived from the measured density jump.

We can predict how the Mach number might change due to calibration differences between NuSTAR and Chandra based on the empirical relation found by F. Lopez et al. (2025, in preparation; Equation (3)):

$$kT_{\rm Nu} = B \cdot kT_{\rm Ch} \cdot e^{-kT_{\rm Ch}/kT_{\rm cut}}, \quad (7)$$

where $kT_{\rm Nu}$ is the predicted NuSTAR temperature, $kT_{\rm Ch}$ is the observed Chandra temperature, and $B = 1.00 \pm 0.05$ and $kT_{\rm cut} = 91 \pm 41$ keV are free parameters determined by the data.

---

[13] https://uncertainties.readthedocs.io/en/latest/#

Since the cross-calibration issue is less prevalent at lower energies, we can be more confident that the pre-shock electron temperature determined from Chandra ($3.2 \pm 1.1$ keV) should be consistent with NuSTAR, although we are unable to directly constrain the temperature in this region. We can then use the Chandra post-shock temperature of $12.0^{+1.4}_{-0.8}$ keV and Equation (7) to predict NuSTAR's post-shock temperature measurement, which yields a temperature of $10.5^{+1.3}_{-1.0}$ keV, a corresponding temperature jump of $3.3 \pm 1.2$, and a Mach number $\mathcal{M} = 2.8 \pm 0.7$. We can compare this to our measured NuSTAR post-shock temperature of $8.7^{+1.8}_{-1.2}$ keV, which, in combination with the Chandra pre-shock temperature, yields a mach number of $\mathcal{M} = 2.5^{+0.7}_{-0.6}$. This however does not include systematic uncertainties that could arise from cross-calibration effects.

At a Chandra temperature of 10 keV, Chandra measures temperatures $15.7\% \pm 4.6\%$ higher than NuSTAR in Chandra's hard band of 3–10 keV (A. N. Wallbank et al. 2022). Since our NuSTAR measurement is ∼15% lower than 10 keV, we can increase our NuSTAR post-shock temperature by $(15.7 \pm 4.6)\%$, to obtain an estimate of what the Chandra temperature should be and the associated Mach number. A $(15.7 \pm 4.6)\%$ increase in our NuSTAR temperature measurement results in an estimated Chandra temperature of $10.1^{+2.1}_{-1.4}$ keV, where the uncertainty of the temperature measurement has been scaled proportionally by the increase and the uncertainty on the increase has been propagated through. This results in a Mach number of $\mathcal{M} = 2.7 \pm 0.7$.

In summary, Chandra measures a post-shock temperature of $12.0^{+1.4}_{-0.8}$ keV and a Mach number of $\mathcal{M} = 3.0 \pm 0.6$. We can estimate what NuSTAR would measure using Equation (7), which yields a NuSTAR post-shock temperature of $10.5^{+1.3}_{-1.0}$ keV and a Mach number of $\mathcal{M} = 2.8 \pm 0.7$. We can also use our actual NuSTAR measurement of the post-shock region ($8.7^{+1.8}_{-1.2}$ keV), which results in a Mach number of $\mathcal{M} = 2.5^{+0.7}_{-0.6}$ or $\mathcal{M} = 2.7 \pm 0.7$, if we increase the temperature by $(15.7 \pm 4.6)\%$ to account for systematic cross-calibration effects. While not entirely consistent, the Mach numbers derived by predicting the NuSTAR temperature from the Chandra measurement and by inferring the Chandra temperature from the NuSTAR measurement (after accounting for systematic cross-calibration effects) are broadly consistent within uncertainties.

Although there are arguments that NuSTAR's temperatures should be more accurate than Chandra's, this has not yet been fully demonstrated; we only know that Chandra and NuSTAR disagree at higher energies, and we can therefore not assume one measurement is more accurate than the other. A comparison of the models derived from each of the Mach numbers discussed in this section is shown in Figure 4.

The collisional model derived using the Mach number $\mathcal{M} = 2.8 \pm 0.7$ is most consistent with our findings, but large uncertainties in our measurements make it impossible to rule out the instant-equilibration model. We can simply say that our results favor collisional equilibration and a Mach number of $\mathcal{M} = 2.8 \pm 0.7$.

*4.2.1. Chandra Temperature Uncertainties*

To make matters more complicated, when a modern analysis of Chandra observations of A665 was carried out, the temperature measurement in the post-shock region next to the shock front did not agree with previous measurements. We measured a projected post-shock temperature of $8.3^{+3.3}_{-2.1}$ keV, while S. Dasadia et al. (2016b) measured a projected post-shock temperature of $11.3^{+2.2}_{-1.8}$ keV. If accurate, a lower post-shock temperature would imply a lower Mach number than previously reported. This was particularly puzzling, as the same data sets were used in both analyses. To reconcile this, we recreated the analysis carried out in S. Dasadia et al. (2016b) step by step. For more details, see Section 2.2.

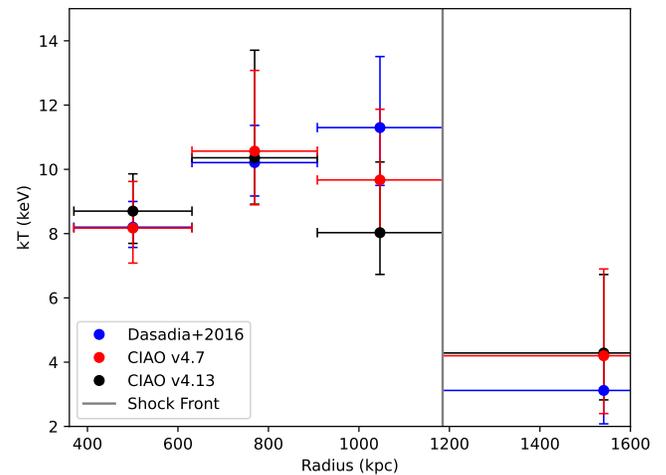

**Figure 5.** A comparison of the projected Chandra temperature measurements in the pre-shock and post-shock regions in A665. The green points correspond to the measurements made in S. Dasadia et al. (2016b), the red points represent an analysis carried out using CIAO v4.7, and the black points represent the measurements using CIAO v4.13. The vertical gray line shows the location of the shock front relative to these regions. All analyses agree with each other with some uncertainty in the immediate post-shock region.

After an attempt to recreate the original analysis, we extracted temperature measurements in identical regions, which mostly agreed with S. Dasadia et al. (2016b). The results can be seen in Figure 5. The post-shock temperature in the region right next to the shock front was still lower than the original measurement in S. Dasadia et al. (2016b), but this can be explained by subtle differences in the final background files used in our analysis. The original analysis used stowed background files for the particle background and included other background components in the spectral modeling. The resulting particle background level from our analysis was slightly higher compared to S. Dasadia et al. (2016b), which could drive the temperature measurement down slightly. The more modern analysis used blank-sky backgrounds with an updated calibration and generated slightly different ARF files, which can explain why the post-shock region's temperature varies across each of the three measurements of the same post-shock region.

When considering the pre-shock region, the original temperature measurements in S. Dasadia et al. (2016b) imply a larger Mach number than would be derived from either of the other sets of our temperature measurements. Ultimately, these differences in the temperature measurements highlight systematic uncertainties in the determination of the Mach number. Since the regions of interest are in the faint parts of the cluster, a more sensitive telescope such as AXIS[14] (C. S. Reynolds et al. 2023) or NewAthena (M. Cruise et al. 2024) will be needed to reconcile the uncertainty in the Mach number.

---

[14] https://axis.umd.edu

### 4.3. In the Context of Previous Work

While electron–ion equilibration models in galaxy cluster mergers were first published in the 1960s (L. Spitzer 1962; Y. Zel'dovich & Y. Raizer 1967), the first measurement and constraint was not made until 2005, in the Bullet Cluster (M. Markevitch 2005), which suggested that instant equilibration is favored in cluster mergers. Since then, a small handful of clusters have had their temperature distribution across their shock mapped and used to constrain the equilibration model. Q. H. S. Wang et al. (2018) looked at a Chandra observation of A520 and found that the temperature directly behind the shock was too high to be attributed to adiabatic compression (as in the delayed-equilibration model), thus also favoring instant equilibration. In contrast, measurements of A2146 led to delayed equilibration being the favored model (H. R. Russell et al. 2012, 2022). When A. Sarkar et al. (2024) mapped the post-shock electron temperature distribution for A1240, their results also favored delayed equilibration. Our NuSTAR measurements of A665 leverage NuSTAR's greater sensitivity at higher energies to constrain temperature measurements of the hot ICM near the shock. We add our measurements and our result of delayed equilibration being favored to the small collection of cluster merger measurements. These contrary results highlight the complexity of cluster mergers and signal the need for more measurements to be made. In particular, electrons that are not in equilibrium systematically bias temperature measurements (C. Avestruz et al. 2015), further exemplifying the need for more merger studies.

### 4.4. Impact on Cosmology

The timescale for equilibration informs what fraction of the kinetic energy in the merger is converted into thermal energy and, in turn, how the gas is thermalized. Since assuming hydrostatic equilibrium (HSE) is necessary to estimate the mass of galaxy clusters from the gas and relies on the assumption of total thermalization, understanding how the gas is thermalized and how the energy is redistributed in the merger is essential to accurately determining cluster masses. We consistently see a discrepancy between the mass function from HSE masses and the mass function expected from $\Lambda$CDM fits to the primary cosmic microwave background anisotropies (A. Vikhlinin et al. 2009; M. Penna-Lima et al. 2017). Since galaxy clusters are the largest gravitationally bound objects in the Universe and lie at the intersections of cosmic filaments, cluster mergers represent the growth of large-scale structure in the Universe. Their masses can be used to set constraints on cosmological parameters. Understanding how the gas is heated in a merger and the timescale for equilibration can help correct HSE mass biases and accurately determine the mass function for galaxy clusters.

## 5. Conclusion

Using NuSTAR observations of A665, we have mapped the electron temperature distribution behind the northern shock front in a powerful galaxy cluster merger, in an attempt to constrain the electron–ion equilibration timescale. We compared two heating models: one where the ions and electrons are immediately heated at the shock front; and one where the electrons are initially adiabatically compressed, then subsequently equilibrate collisionally with the ions over $\sim$100 Myr. Our measurements of the post-shock temperature distribution in A665 favor the collisional model and a Mach number of $\mathcal{M} = 2.8 \pm 0.7$. We determine the electron–ion equilibration timescale in A665 to be $t_{eq} = (4.0 \pm 3.4) \times 10^8$ yr and add our result to the small handful of merging clusters that have had their post-shock electron temperatures mapped. Our constraint of favored delayed equilibration aligns with the results for A1240 (A. Sarkar et al. 2024) and A2146 (H. R. Russell et al. 2012, 2022), but is in contrast to the instant-model constraints for A520 (Q. H. S. Wang et al. 2018) and the Bullet Cluster (M. Markevitch 2005). Identifying the heating scenario gives a timescale for the thermalization of the ICM in a galaxy cluster merger, and since calculating the mass of a galaxy cluster relies on the assumption of fully thermalized gas, understanding the timescale for thermalization is key to accurately determining cluster masses and using those estimates to constrain cosmology. Our measurements are affected by various systematics, including those present in spectral fitting and projection effects. Since the regions of interest are in the fainter parts of the cluster, deeper observations and a more sensitive instrument will be needed to effectively combat systematics for future studies.

## Acknowledgments

D.R.W. acknowledges support from the NASA NuSTAR grant NNX17AH31G. C.L.S. was supported in part by NASA XMM-Newton grants 80NSSC22K1510 and 80NSSC24K1518 to the University of Virginia, and by the Virginia Institute for Theoretical Astrophysics (VITA), supported by the College and Graduate School of Arts and Sciences at the University of Virginia.

Special thanks to the CIAO and HEASOFT help desks for assisting in the retrieval and setting up of outdated files and software.

# Appendix A
# Processing and Analysis Details

## A.1. Background Fitting

Figure 6 is a detailed figure displaying each NuSTAR observation in Table 1 for each NuSTAR detector, with background regions overlaid, compared to the background spectra extracted and fit from each region, as described in Section 2.

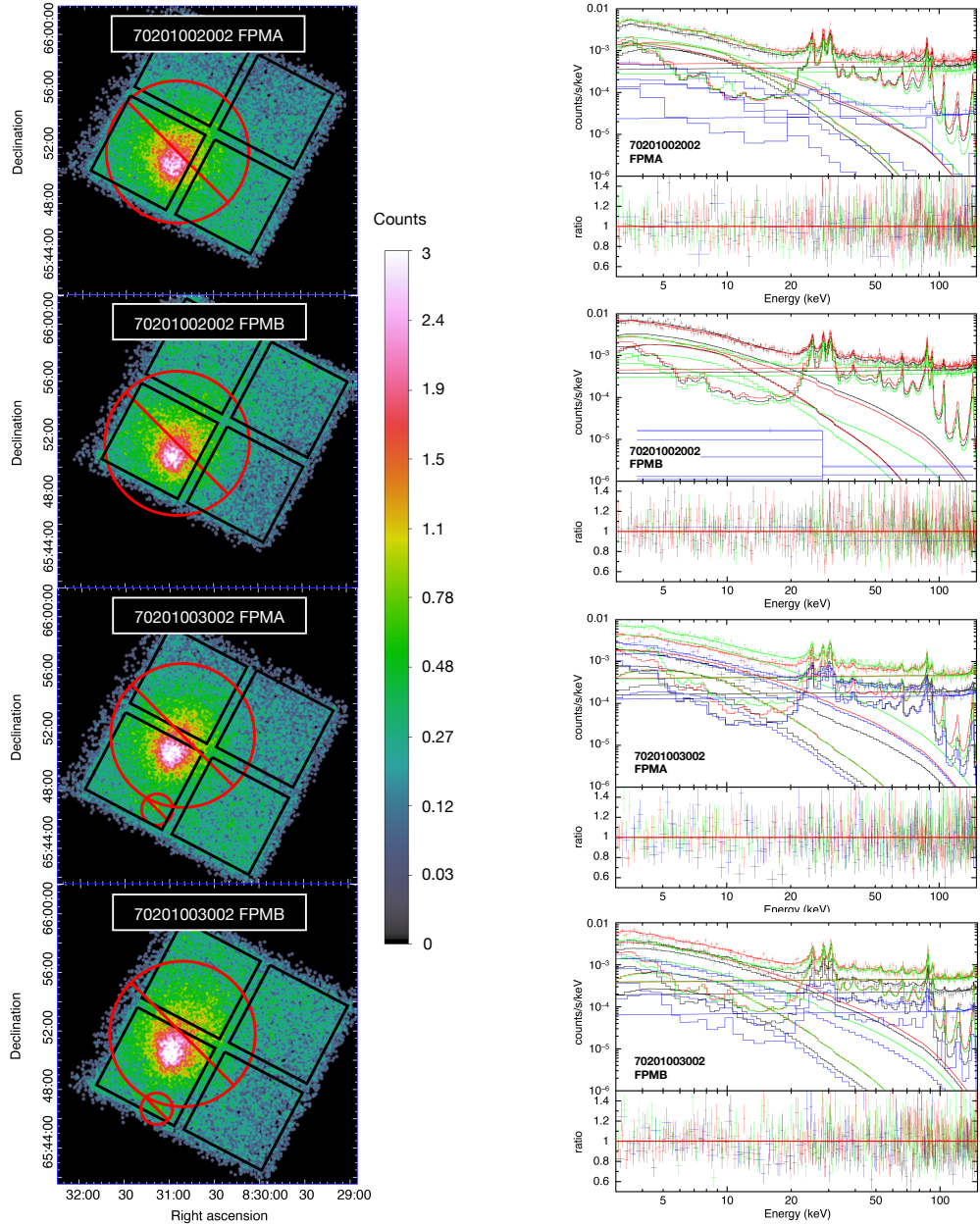

**Figure 6.** Left: background regions (black) used in the nuskybgd background analysis in Section 2, overlaid on raw count images (smoothed with a Gaussian kernel with an FWHM of 3 pixels) of A665, with the cluster emission exclusion region in red. The top two panels are for OBSID 70201002002, and the bottom two panels are for OBSID 70201003002. Within each pair, the top image corresponds to NuSTAR's FPMA detector and the bottom image to the FPMB detector. The color bar is log scaled and represents raw counts. Right: joint fits of the background spectra extracted from NuSTAR's FPMA and FPMB for both observations (OBSID 70201002002 for the top two panels, and OBSID 70201003002 for the bottom two panels). The individual colors correspond to each background region used in the extraction and fitting. The spectral bins are grouped to have a detection of at least $8\sigma$, with a maximum of 12 bins.

*A.2. Initial Temperature Map*

The initial temperature map, used as a reference in our analysis, as described in Section 3.1, can be found below in Figure 7.

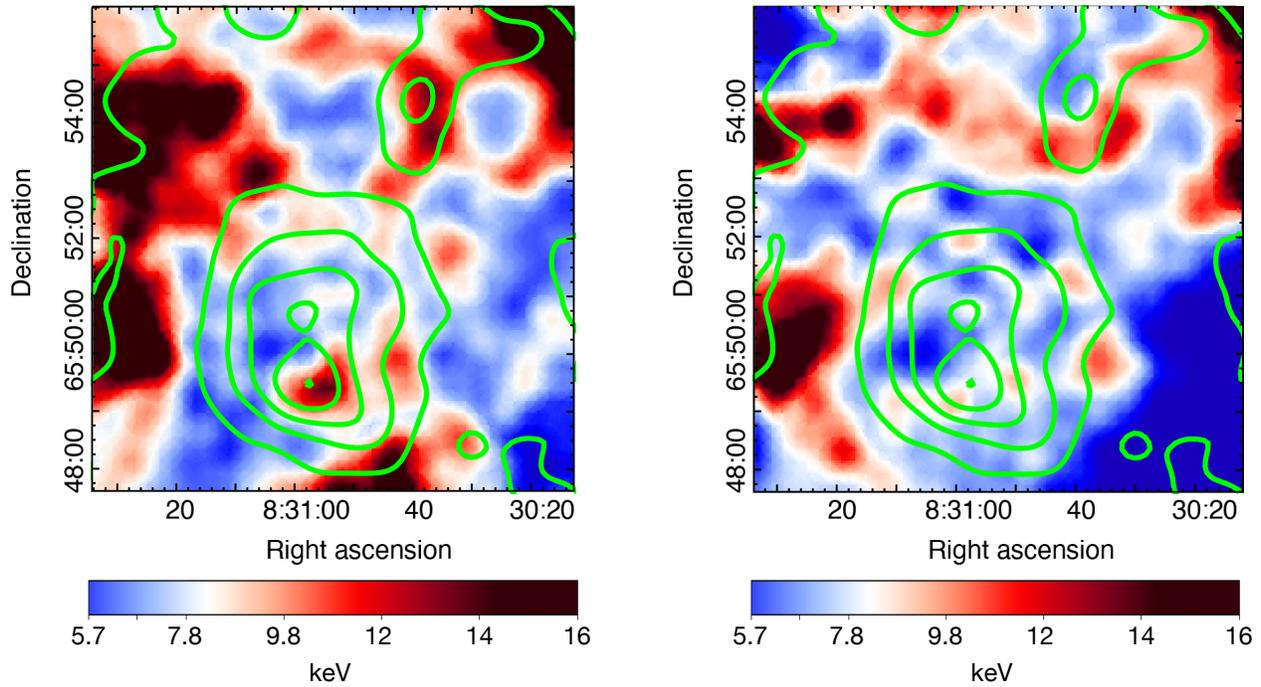

**Figure 7.** A rough temperature map of A665 (left: OBSID 70201002002; right: OBSID 70201003002) calculated through the spectral fitting of predefined pixel areas, as described in Section 3.1. The green contours outline the LOFAR radio emission in Figure 3, matching the contours shown in Figure 1. The color bar indicates the temperature in keV.

### A.3. Accounting for NuSTAR's PSF

Figure 8 demonstrates the quality of our fits in the nuCrossARF analysis, described in Section 3.3.

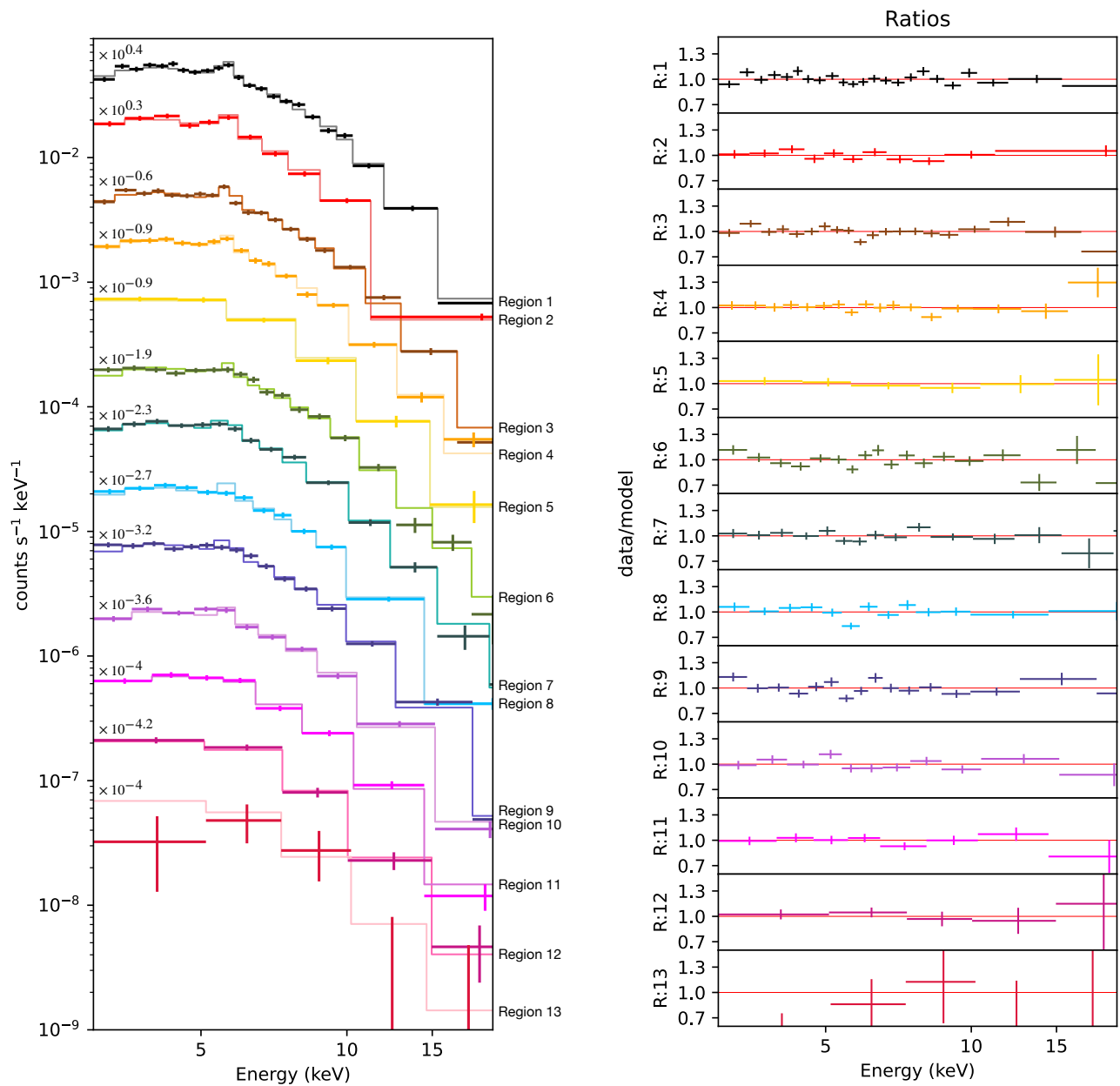

**Figure 8.** Left: joint spectral fits for the nuCrossARF analysis (Section 3.3) using the regions shown in Figure 2. A combined model accounting for the contribution from each region for each observation (solid) is compared to spectral data combined from each of the four observations (points). This is purely for illustrative purposes, as each observation was fit individually. Each set of data and a corresponding model is shifted vertically on a logarithmic scale for illustrative purposes ($\times 10^{shift}$). Right: the ratio showing the goodness of the fit for each region. The region numbers are listed on the y-axis, and the colors correspond with the data and models on the left.

## Appendix B
## Collisional Electron–Ion Equilibration

The equilibration timescale between two particles depends on each particle's mass ($m$), charge ($Ze$), and Coulomb Logarithim ($\ln\Lambda$)—e.g., Equations (5)–(31) from L. Spitzer (1962):

$$t_{\rm eq} = \frac{3 m m_f k^{3/2}}{8(2\pi)^{1/2} n_f Z^2 Z_f^2 \ln\Lambda} \left(\frac{T}{m} + \frac{T_f}{m_f}\right)^{3/2}, \quad (B1)$$

$$= 1.87 \times 10^8 \, {\rm yr} \, \frac{A_f}{Z^2 Z_f^2 A^{1/2}} (\ln\Lambda)^{-1}$$
$$\times \left(1 + \frac{T_f}{T}\frac{A}{A_f}\right)^{3/2} \left(\frac{T}{10^8 {\rm K}}\right)^{3/2} \left(\frac{n_f}{10^{-3}{\rm cm}^{-3}}\right)^{-1}, \quad (B2)$$

where $A = m/m_p$, the mass of the particle scaled by the proton mass; $Z$ is the atomic number; and $T$ is the temperature of the gas. The subscript "$f$" represents the field particle (the ions). The relevant Coulomb logarithm for electrons for temperatures $T_e \geqslant 4 \times 10^5 K \sim 0.0347$ keV is (from C. Sarazin (1988); Equation (5.33)):

$$\ln\Lambda = 37.8 + \ln\left[\left(\frac{T}{10^8 \ {\rm K}}\right)\left(\frac{n_e}{10^{-3} \ {\rm cm}^{-3}}\right)^{-1/2}\right]. \quad (B3)$$

For electrons and ions, the collisional equilibration timescale is thus

$$t_{\rm eq} \approx 2.33 \times 10^8 \left(\frac{T_e}{10^8 \ {\rm K}}\right)^{3/2}$$
$$\times \left(\frac{n_e}{10^{-3} \ {\rm cm}^{-3}}\right)^{-1} \, {\rm yr}, \quad (B4)$$

where $T_e/T_i \gg m_e/m_p$.

The energy exchange between the electrons and ions is given by L. Spitzer (1962; Equations (5)–(30)):

$$\frac{dT_e}{dt} = \frac{T_i - T_e}{t_{eq}}. \quad (B5)$$

The differential equation can be solved analytically, to find the evolution of the electron temperature through the shock.

We follow the same approach as Appendix C of C. L. Sarazin et al. (2016) and define the terms more explicitly, including a dimensionless parameter $\tau = T_e/T_g$. The energy exchange can then be rewritten as

$$\frac{d\tau}{dt} = 2.8 \times 10^{-1} \ {\rm s}^{-1} \left(\frac{T_g}{\rm K}\right)^{-3/2}$$
$$\times \left(\frac{n_{e,{\rm post}}}{\rm cm^{-3}}\right) \frac{n_e}{n_{e,{\rm post}}} \frac{1-\tau}{\tau^{3/2}}. \quad (B6)$$

We define the density as a power law: $n_e(r) \propto r^{-p}$. The post-shock gas velocity can also be included as $v_{\rm ps} = \frac{dr}{dt}$, where the post-shock gas velocity can be considered as constant and is determined by dividing the shock speed by the density jump at the shock, where the density jump is directly related to the Mach number by Equation (6): $v_{\rm ps} = v_{\rm shock}/\left(\frac{\rho_{\rm post}}{\rho_{\rm pre}}\right)$, where the shock speed is determined by multiplying the Mach number by the pre-shock sound speed: $v_{\rm shock} = \mathcal{M} \cdot c_s$.

The differential equation can then be rewritten as

$$\frac{d\tau}{dx} = C \cdot x^{-p} \cdot \frac{1-\tau}{\tau^{3/2}}, \quad (B7)$$

where $x = r/r_{\rm sf}$, $r_{\rm sf}$ is the radius of the shock front, and $C = 9.8 \times 10^{15} \left(\frac{r_{\rm sf}}{\rm kpc}\right)\left(\frac{v_{\rm ps}}{\rm km} {\rm s}^{-1}\right)^{-1} \left(\frac{T_g}{K}\right)^{-3/2} \left(\frac{n_{e,{\rm post}}}{\rm cm}^{-3}\right)$. The differential equation can then be solved:

$$\left[-\frac{2}{3}\tau^{3/2} - 2\tau^{1/2} + \ln\left(\frac{\tau^{1/2}+1}{|\tau^{1/2}-1|}\right)\right]_{\tau_0}^{\tau}$$
$$= C \times \begin{cases} \frac{1}{1-p}(x^{1-p} - 1), & p \neq 1 \\ \ln(x), & p = 1 \end{cases}, \quad (B8)$$

where $\tau_0 = T_{e,{\rm post}}/T_g$, and the post-shock electron temperature $T_{e,{\rm post}}$ is calculated from adiabatic compression at the shock front (Equation (2)).

It is difficult to solve for the electron temperature as a function of distance, but we can instead solve for the distance as a function of temperature, yielding the desired result:

$$r = r_{\rm sf} \times \begin{cases} \left\{\left[-\frac{2}{3}\tau^{3/2} - 2\tau^{1/2} + \ln\left(\frac{\tau^{1/2}+1}{|\tau^{1/2}-1|}\right)\right]_{\tau_0}^{\tau} \frac{p-1}{C} + 1\right\}^{1/(1-p)}, & p \neq 1 \\ \exp\left\{-\frac{1}{C}\left[-\frac{2}{3}\tau^{3/2} - 2\tau^{1/2} + \ln\left(\frac{\tau^{1/2}+1}{|\tau^{1/2}-1|}\right)\right]_{\tau_0}^{\tau}\right\}, & p = 1 \end{cases}. \quad (B9)$$

### ORCID iDs

Christian T. Norseth ⓘ https://orcid.org/0000-0001-9389-6050
Daniel R. Wik ⓘ https://orcid.org/0000-0001-9110-2245
Craig L. Sarazin ⓘ https://orcid.org/0000-0003-0167-0981
Ming Sun ⓘ https://orcid.org/0000-0001-5880-0703
Fabio Gastaldello ⓘ https://orcid.org/0000-0002-9112-0184